\documentclass[bibyear]{aa}  

\usepackage{graphicx}
\usepackage{txfonts}
\begin{document}

   \title{Neighbours hiding in the Galactic plane, a new M/L dwarf 
          candidate for the 
          8~pc sample
           }


\titlerunning{New M/L dwarf candidate for the 8~pc sample}

   \author{R.-D. Scholz}

   \institute{Leibniz-Institut f\"ur Astrophysik Potsdam (AIP),
              An der Sternwarte 16, 14482 Potsdam, Germany\\
              \email{rdscholz@aip.de}
             }

   \date{Received 8 November 2013; accepted 12 December 2013}

 
  \abstract
   {}
   {Using Wide-field Infrared Survey Explorer (WISE) data and previous 
    optical and near-infrared sky surveys, we try to
    identify still missing stellar and substellar neighbours of the Sun.}
   {When checking the brightest red WISE sources for proper motions and colours
    expected for nearby M and L dwarfs, we also approached the thin Galactic 
    plane. Astrometry (proper motion and parallax measurements) and the 
    available photometry were used to
    obtain first estimates of the distance and type of nearby candidates.}
   {We have discovered WISE~J072003.20$-$084651.2, an object with moderately 
    high proper motion ($\mu$$\approx$120~mas/yr) that lies at low Galactic
    latitude ($b$$=$$+$2.3$\degr$), with similar
    brightness ($J$$\approx$10.6, $w2$$\approx$8.9) and colours 
    ($I$$-$$J$$\approx$3.2, $J$$-$$K_s$$\approx$1.2, $w1$$-$$w2$$\approx$0.3)
    as the nearest known M-type brown dwarf LP~944-20. With a photometric
    classification as an M9$\pm$1 dwarf, its photometric
    distance lies in the range between about 5 and 7~pc, based on
    comparison with absolute magnitudes of LP~944-20 alone or
    of a sample of M8-L0 dwarfs. The slightly larger 
    distance derived from our preliminary trigonometric parallax 
    (7.0$\pm$1.9~pc) may indicate a close binary nature. The new neighbour
    is an excellent target for planet search and low-mass star/brown
    dwarf studies.} 
   {}

   \keywords{
Astrometry --
Proper motions --
Stars: distances --
Stars:  kinematics and dynamics  --
brown dwarfs --
solar neighbourhood
               }

   \maketitle


\section{Introduction}
\label{Sect_intro}

The incomplete knowledge of our nearest stellar neighbours, demonstrated, e.g., 
by Henry et al.~(\cite{henry97}), has motivated many astronomers 
to take part in the search for still missing late-type dwarfs and
white dwarfs surrounding the Sun. The 74 AFGK stars within 10~pc have already 
been known for a long time and their trigonometric parallaxes were measured 
by Hipparcos (ESA~\cite{esa97}; van Leeuwen~\cite{vanleeuwen07}). In contrast,
the numbers of M dwarfs and white dwarfs increased from 198 to 248 and
from 18 to 20, respectively, during the last 12 years, according to the
Research Consortium on Nearby Stars 
(RECONS)\footnote{http://www.chara.gsu.edu/RECONS/}. This progress
was achieved thanks to many individual discoveries and subsequent 
time-consuming trigonometric parallax measurements of the new candidates.
The RECONS error requirement for the parallax of less than 10~mas means
that the distance for every RECONS 10~pc sample member is known to better than 
10\%. As of 1 January 2012, there are also five L and ten T dwarfs falling in the 
RECONS 10~pc sample. Only the nearest 100 systems ($d$$<$6.7~pc) of this 
high-quality sample are published on the RECONS Web site.

An updated 8~pc sample, which includes objects with less accurate 
and photometric distance estimates, was compiled by 
Kirkpatrick et al.~(\cite{kirkpatrick12}). This sample was used to explore 
the field brown dwarf mass function, but most of the 22 T- and 8 Y-type brown
dwarfs listed still need higher accuracy to meet the RECONS requirement. Only
three L dwarfs are known within 8~pc, and all these are late-L types. Late-M
and L dwarfs may be stellar or substellar depending on their age, and
lithium absorption can be used as a distinguishing criterion (Rebolo, 
Mart{\'{\i}}n, \& Magazzu~\cite{rebolo92}).
Among six late-M ($>$M8) dwarfs in the 8~pc sample of 
Kirkpatrick et al.~(\cite{kirkpatrick12}), there is one M9 dwarf (LP~944-20)
that was correspondingly classified as a relatively young (a few hundred Myr
compared to several Gyr typical for the solar
neighbourhood) brown dwarf (Tinney~\cite{tinney98}, Ribas~\cite{ribas03},
Pavlenko et al.~\cite{pavlenko07}).

The Galactic plane, a region often excluded or considered
as problematic in previous searches due to image crowding, is probably
still hiding many close neighbours, mainly cool stars, and even cooler
brown dwarfs. 
Some recent large-area searches successfully
approached (Looper et al.~\cite{looper08}) or included
(Phan-Bao et al.~\cite{phanbao08},
Folkes et al.~\cite{folkes12},
Luhman~\cite{luhman13}) the Galactic plane. 
Dedicated Galactic plane surveys 
(Lucas et al.~\cite{lucas10},
Burningham et al.\cite{burningham11},
Beam{\'{\i}}n et al.~\cite{beamin13})
have also succeeded in finding new neighbours.

Here, we introduce a previously missing probable 8~pc sample member in the 
Galactic plane, WISE~J072003.20$-$084651.2 (hereafter WISE~J0720$-$0846),
which was identified among bright candidates selected from Wide-field 
Infrared Survey Explorer (WISE; Wright et al.~\cite{wright10}) data 
(Sect.~\ref{Sect_search}). Despite its relatively small proper motion 
(Sect.~\ref{Sect_pmplx}), the preliminary trigonometric parallax, based
on the available 12 epochs of observations
(Sect.~\ref{Sect_pmplx}), leads to a distance of about 7~pc. 
Its magnitudes
and colours, which are very similar to those of the M9 dwarf LP~944-20, 
place it at a photometric distance of about 5-7~pc
(Sect.~\ref{Sect_phot}).

   \begin{figure*}
   \centering
   \includegraphics[width=13.8cm]{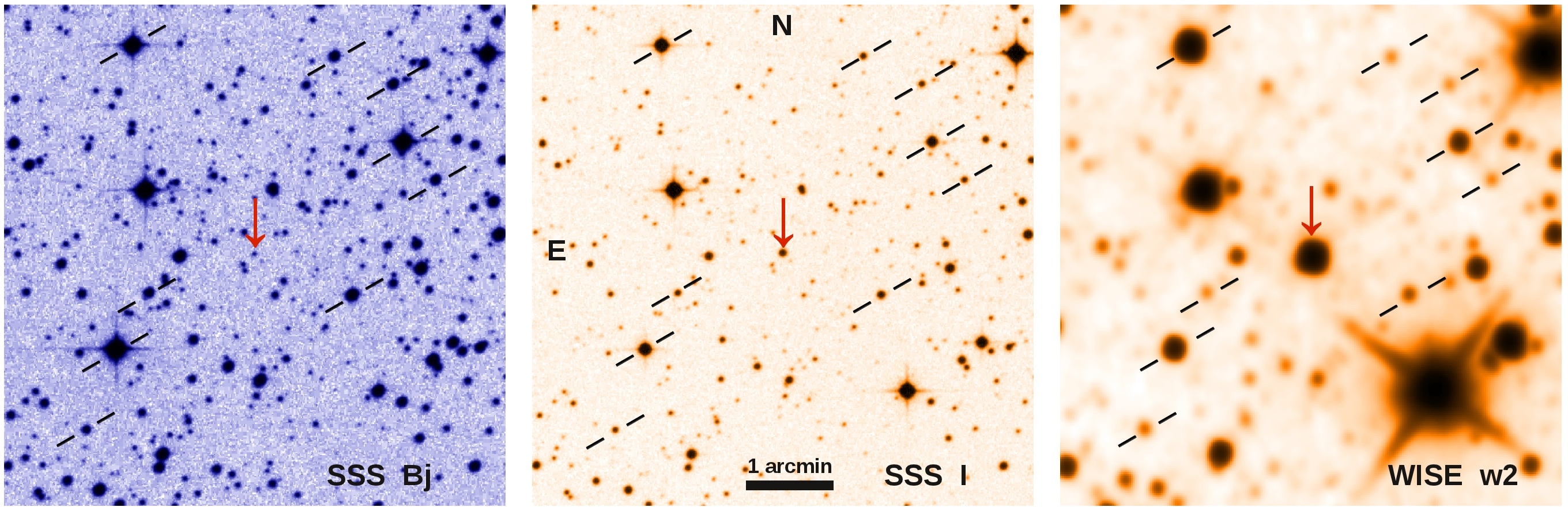}
      \caption{Finder charts of 6$\times$6~arcmin$^2$ centred
              on WISE~J0720$-$0846 from
              SuperCOSMOS Sky Surveys (SSS) of $B_J$-band (left)
              and $I$-band (centre) photographic plates, and
              from WISE $w2$-band (right) observations. The target
              is marked by an arrow, and nine reference stars used
              for positional corrections (Sect.~\ref{Sect_pmplx})
              are marked by dashes, respectively.
              }
         \label{fig1_3images}
   \end{figure*}

\begin{table*}
\caption{Multi-epoch positions, their uncertainties, applied corrections (offsets), and photometry of WISE~J0720$-$0846.}
\label{table_pos}      
\centering                          
\begin{tabular}{llrrrclll}        
\hline\hline                 
$\alpha$ (J2000) & $\delta$ (J2000) & $\sigma_{\alpha,\delta}$  & $\Delta_{\alpha}$ & $\Delta_{\delta}$ & Epoch    & Source & Photometry\tablefootmark{b} \\
($\degr$) & ($\degr$) & (mas) & (mas) & (mas) & (year) & & (mag) \\
\hline                        
110.014078 & $-$8.779181 & 240 &  $+$20 & $-$307 & 1955.882 & SSS $E$ & $E_{pg}$$=$17.08 \\
110.014088 & $-$8.779175 & 370 &  $+$78 & $+$351 & 1955.882 & DSS $O$\tablefootmark{a} \\
110.013690 & $-$8.779973 &  70 &  $+$27 &  $+$85 & 1981.188 & SSS $I$ & $I_{pg}$$=$12.91 \\
110.013653 & $-$8.780144 & 220 &  $-$16 & $+$119 & 1985.953 & SSS $R$ & $R_{pg}$$=$16.80 \\
110.013642 & $-$8.780193 & 350 &  $+$57 &  $-$51 & 1990.971 & SSS $B_J$&$B_J$$_{pg}$$=$19.76 \\
110.013581 & $-$8.780536 &  50 &  $-$85 & $+$184 & 1998.978 & DENIS & $I$$=$13.805 $J$$=$10.674 $K_s$$=$9.399 \\
110.013557 & $-$8.780529 &  50 &      0 &      0 & 1999.142 & 2MASS & $J$$=$10.628 $H$$=$9.919 $K_s$$=$9.467 \\
110.013454 & $-$8.780566 & 400 & $+$120 &  $-$70 & 2001.005 & SSS H$_{\alpha}$ & $RHa$$_{pg}$$=$16.11 \\
110.013400 & $-$8.780524 & 500 &  $+$51 &  $-$25 & 2001.049 & SSS short-R & $SR2$$_{pg}$$=$17.00 \\
110.013484 & $-$8.780679 & 150 &  $+$28 &  $+$17 & 2004.066 & CMC & $r'$$=$16.850 \\
110.013350 & $-$8.780903 &  60 &   $-$6 &   $-$2 & 2010.266 & WISE all-sky& $w1$$=$9.169 $w2$$=$8.857 $w3$$=$8.278 \\
110.013415 & $-$8.780936 & 120 &  $-$11 &  $-$29 & 2010.792 & WISE post-cryo\tablefootmark{c} & \\
\hline                        
\end{tabular}
\tablefoot{The $\alpha,\delta$ positions of the target are already 
corrected for the mean offsets ($\Delta_{\alpha}$, $\Delta_{\delta}$) of the 
reference stars with respect to their 2MASS positions.
\tablefoottext{a}{Position measured on FITS image downloaded from http://archive.stsci.edu/cgi-bin/dss\_plate\_finder.}
\tablefoottext{b}{Photographic magnitudes marked as $_{pg}$ are accurate to about 0.1-0.2~mag. Other magnitudes listed are accurate to about 0.02~mag, except for DENIS $J$ ($\pm$0.06~mag) and $K_s$ ($\pm$0.09~mag), and the CCD-based CMC $r'$ (unknown).}
\tablefoottext{c}{Averaged coordinates from 14 single exposures.}
}
\end{table*}


\section{Candidate search and cross-identification}
\label{Sect_search}

Our search strategy aimed at hidden M- and L-type low-mass
stellar and substellar neighbours of the Sun. They appear very bright in 
the mid-infrared WISE survey and can be separated from distant reddened 
stars and red giants by their characteristic
optical-to-infrared colours and high proper motions (HPMs) 
(see, e.g., Phan-Bao et al.~\cite{phanbao08}). To measure these
properties, a cross-identification with their correct counterparts
in near-infrared and optical surveys is necessary. Our idea was
to check this cross-identification for every bright WISE source 
with WISE colours expected for dwarfs of spectral type $\ge$M5
(see Kirkpatrick et al.~\cite{kirkpatrick11}). This is, therefore, a
combined HPM and colour-based search restricted to
the brightest and hence nearest M/L dwarfs. In comparison to other
HPM surveys such as, e.g., 
SUPERBLINK (started by L{\'e}pine, Shara \& Rich~\cite{lepine02}) and
SuperCOSMOS-RECONS (started by Hambly et al.~\cite{hambly04}),
this is a small survey (or completeness check)
dedicated to a certain class of objects in the solar neighbourhood.

To select nearby M/L dwarf candidates,
we applied the following selection 
criteria to the WISE All-Sky source catalogue:

\begin{enumerate}
\item $w1$$-$$w2$$\ge$0.2 and $w2$$<$10
(compare with Figs.1 and 29 in Kirkpatrick et al.~\cite{kirkpatrick11})
\item $w2$$-$$w3$$<$2.5 and $w3$$-$$w4$$<$2.0 
(similar colour cuts were described by Luhman~\cite{luhman13}, for excluding
extragalactic sources)
\end{enumerate}

All WISE sources (including point and extended sources)
were considered as M/L dwarf candidates if they matched the criteria
listed above. After pre-selecting the M/L dwarf candidates,we 
performed two different searches for nearby HPM objects among them
using the cross-identification with the
Two Micron All Sky Survey (2MASS; Skrutskie et al.~\cite{skrutskie06})
listed in the WISE All-Sky source catalogue. 

First (3a), a WISE
M/L candidate was flagged as a HPM candidate if there was no 2MASS source 
within 3~arcsec of this WISE candidate. This all-sky search was sensitive 
to all red WISE objects brighter than $w2$$=$10 with proper motions larger
than about 300~mas/yr because the typical epoch difference between 2MASS
and WISE is about ten years.

In our second search (3b), we focused on stars with $|b|$$>$2$\degr$
to exclude the very thin Galactic plane and considered all WISE
M/L candidates as possible HPM objects if there was a 2MASS source
between 1~arcsec and 3~arcsec radius of their WISE position. This 
search was sensitive to stars with proper motions between about
100 and 300~mas/yr.

   \begin{figure}
   \centering
   \includegraphics[width=5.2cm]{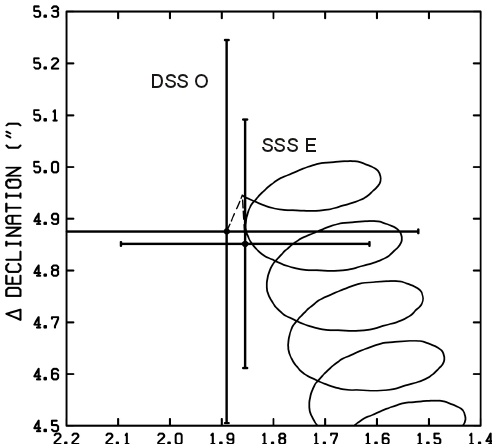}
   \includegraphics[width=5.2cm]{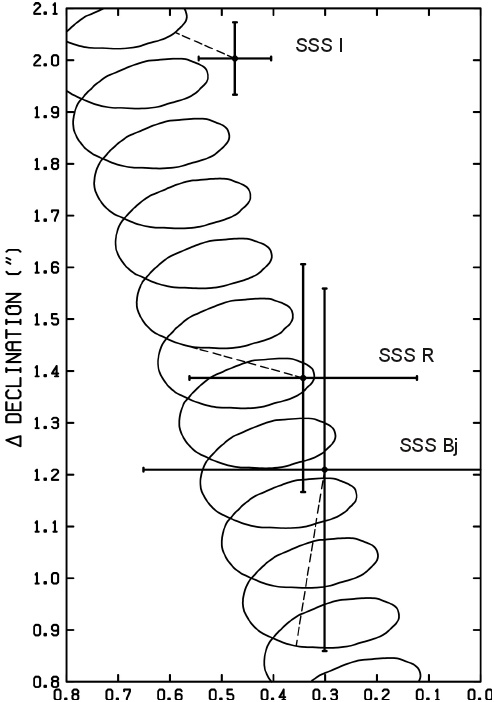}
   \includegraphics[width=5.2cm]{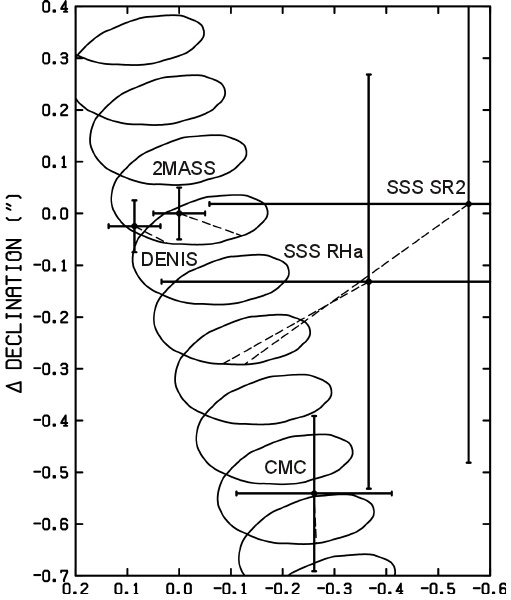}
   \includegraphics[width=5.2cm]{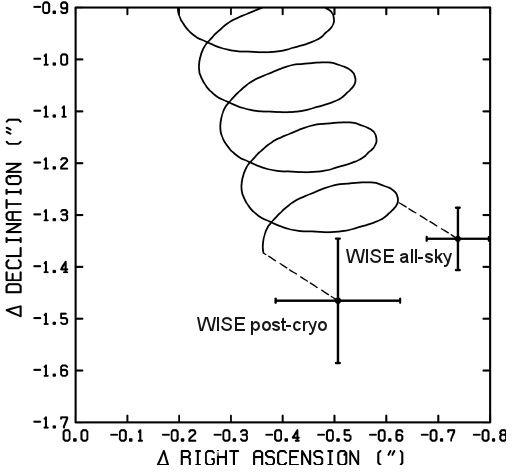}
      \caption{Relative astrometry of WISE~J0720$-$0846 after conversion of
               the positions from DSS, SSS, DENIS, CMC, and WISE to 
               a common system represented by 2MASS
               (points with error bars; see Table~\ref{table_pos} 
               and text) and the combined
               proper motion and parallax solution (curve, 
               see Table~\ref{table_pmplx}). Dashed lines
               show the displacements of observed positions 
               relative to the fit. The panels show (from top to bottom)
               observations from 1955, 1981-1990, 1998-2004, and 2010,
               respectively.}
         \label{fig2_plxpm}
   \end{figure}

The resulting numbers of candidates, without and with 2MASS counterparts
in the WISE catalogue (search (3a) and (3b), respectively),
were in each case about 9500. These two samples were then
cross-identified with known SIMBAD objects within a search radius of
60~arcsec using the cross-match service provided by the 
Centre de Donn{\'e}es astronomiques de Strasbourg (CDS). All
candidates not identified with previously known nearby 
or HPM objects were cross-checked in the Dwarf\-Archives in addition 
(Gelino, Kirkpatrick \& Burgasser~\cite{gelino12}), and, also
visually inspected with the help of the IRSA Finder Charts
tool\footnote{http://irsa.ipac.caltech.edu/applications/finderchart/}
providing Digitised Sky Surveys (DSS), 2MASS, and WISE
images for a given object at a glance. 

Among the candidates  of search (3a), 
the majority turned out to be false HPM detections, whereas
we found about 1000 known HPM, but only about ten new HPM objects.
The latter will be published elsewhere. Their estimated photometric
distances lie in the range between 10 and 30~pc.
The nearest known
objects from the 8~pc sample of Kirkpatrick et al.~(\cite{kirkpatrick12}) were
used for checking the success rate of this first search. Among the late-M
($\ge$M5) dwarfs in the 8~pc sample, there are 44 that have no close 
companions of earlier spectral type and proper motions larger than
300~mas/yr. Six of them have 0.15$<$$w1$$-$$w2$$<$0.2,
as can be expected due to the spread seen in 
Fig.~1 of Kirkpatrick et al.~(\cite{kirkpatrick11}), whereas two have 
even negative colour indices due to an overlap with background stars.
The remaining 36 objects were included in our first candidate list and
also re-discovered as HPM objects so that the success
rate for the late-M dwarfs of the 8~pc sample was 82\%. All three
late-L dwarfs in the 8~pc sample (L5, L8, and L9) are fainter 
(10$<$$w2$$<$11) than the survey limit set by condition 1.

The success rate for the
second search (3b), for moderately HPM 
objects, is difficult to estimate. The reasons are
a) there are only three known $\ge$M5 dwarfs (without close
companions of earlier spectral type) with proper motions below
300~mas/yr in the 8~pc sample, and b) this second search is still
ongoing. However, it already
led to an intriguing new bright red
($J$$=$10.6, $J$$-$$K_s$$=$1.2, $J$$-$$w2$$=$1.8)
candidate at very low Galactic latitude
($b$$=$$+$2.3$\degr$). Its proper motion was clearly visible as the optically
fainter
counterpart on the red DSS1 ($E$) plate was shifted by a few arcseconds
with respect to the 2MASS position. We
extracted the multi-epoch positions measured on four plates ($E$, $B_J$, $R$,
and $I$) within the SuperCOSMOS Sky Surveys
(SSS; Hambly et al.~\cite{hambly01}) together with two more positions
($RHa$ and $SR2$)
measured in the SuperCOSMOS H$_{\alpha}$ survey (Parker et al.~\cite{parker05}).
The object was also detected with a large colour index $I$$-$$J$$\approx$3.2
in the DEep Near-Infrared Survey
(DENIS; Epchtein et al.~\cite{epchtein97}), and we extracted 
positional measurements (and an $r'$ magnitude) from 
the Carlsberg Meridian Catalog (CMC) (\cite{cmc06}) (Table~\ref{table_pos})
that helped to increase the parallax factor coverage.


\section{Proper motion and parallax}
\label{Sect_pmplx}

Proper motion measurements of WISE~J0720$-$0846 were already listed
in the USNO B1.0 (Monet et al.~\cite{monet03}) and PPMXL (R\"oser,
Demleitner \& Schilbach~\cite{roeser10}) catalogues. The B1.0 catalogue lists
($-$36, $-$102)$\pm$(13, 27)~mas/yr, whereas the PPMXL values are
more precise: ($-$39, $-$107)$\pm$(7, 7)~mas/yr. From the collected 12
epochs (Table~\ref{table_pos}), We expected to further improve the proper motion
and also aimed at a preliminary parallax measurement. For this purpuse,
we used the software of Gudehus~(\cite{gudehus01})
allowing for a combined proper motion and parallax solution with
weights corresponding to the positional errors of the different surveys.
Before that, we computed the mean offsets of some reference stars around
the target with respect to their 2MASS coordinates
to bring the different positional measurements of the target
to a common system, A similar procedure was also applied
by Luhman~(\cite{luhman13}) before he determined the parallax and proper
motion of the very close ($d_{trig}$$=$2.0$\pm$0.15~pc) new neighbour
of the Sun, WISE~J104915.57$-$531906.1, using its DSS, DENIS, 2MASS, and 
WISE positions.

Suitable reference stars were selected within 3~arcmin of the 2MASS position 
of WISE~J0720$-$0846 if they:
\begin{itemize}
\item appeared as isolated, well-measured objects
on/in each of the 12 plates/surveys,
\item had high-quality flags (AAA) and $J$$<$16 in 2MASS,
\item had no significant proper motion in the PPMXL ($<$10~mas/yr), and
\item were not too bright or too faint in the WISE post-cryo data
(large positional spread of their inividual exposures).
\end{itemize}
With these conditions we were left with nine reference stars 
(Fig.~\ref{fig1_3images}), out of
17 possible stars before applying the last criterion. 
This figure also shows the changes in resolution
and brightness of the stars with the different passbands.
All coordinates given in Table~\ref{table_pos} are already corrected for
the mean offsets $\Delta$
of the reference stars, also shown in the table. Some of these offsets
were relatively large, as one can see, e.g., for SSS $E$ and DSS $O$, but
also for DENIS. The originally measured declination values on the SSS $E$ and
DSS $O$ plates were strongly deviating, but the corrected values are in good
agreement.

The standard deviations of the positional differences of the
reference stars vary between 20 and 80~mas for DENIS, CMC, SSS $I$, and WISE,
whereas for the remaining red and blue photographic plate measurements from
SSS and DSS the standard deviations increase to 100-250~mas.
As the latter values are much larger than the effect of
differential colour refraction (Deacon et al.~\cite{deacon05}),
we did not attempt to correct the target positions for these systematic errors.
The uncertainties $\sigma_{\alpha,\delta}$ for the
multi-epoch target positions were determined based on the
described standard deviations obtained from the reference stars,
but they were not allowed to fall short of a lower limit of 
50~mas for 2MASS. For the optical observations,
we assumed larger errors for the position of the target
compared to the standard deviations of the reference stars
if the target appeared very faint or distorted (elliptical).
Although the CMC-2MASS standard deviations of the 
reference stars were among the smallest, confirming the
good astrometric quality of the CMC, we adopted larger 
$\sigma_{\alpha,\delta}$ for the CMC position of the target with a
magnitude close to the CMC limit at $r'$$\approx$17. 
Elliptical images in the case of SSS H$_{\alpha}$ and
SSS short-R were the reason for assigning the largest $\sigma_{\alpha,\delta}$
to these measurements. The chosen $\sigma_{\alpha,\delta}$ for the WISE
all-sky and post-cryo, 2MASS, DENIS, and SSS $R$ positions are
comparable to those used by Luhman~(\cite{luhman13}). He only adopted
higher values due to a blended image of his target in the case of his
DSS $I$ position.

The derived proper motion and parallax is provided in Table~\ref{table_pmplx}.
The (relatively poor) quality of the combined fit  
is illustrated in Fig.~\ref{fig2_plxpm},
where the displacements of the observed positions relative to the fit
are shown.
The parallax error is of the same order as that derived by 
Luhman~(\cite{luhman13}) with input data of similar quality.
However, it is larger than the error measured by Scholz et al.~(\cite{scholz12})
for another new neighbour, WISE~J025409.45$+$022359.1, for which more
accurate input data and more epochs were available. Due to the smaller
parallax of WISE~J0720$-$0846 the error leads to a much larger relative 
error in the distance (27\% instead of 7.5\% for Luhman's object).
This trigonometric parallax can only be considered as a very preliminary 
result as it was obtained from very different surveys.

\begin{table}
\caption{Proper motion, parallax, distances, and  tangential velocity.} 
\label{table_pmplx}      
\centering                          
\begin{tabular}{lrl}        
\hline\hline                 
Parameter & Value & Unit \\
\hline
$\mu_{\alpha}\cos{\delta}$ & $-$0.041$\pm$0.002 & arcsec/yr \\
$\mu_{\delta}$             & $-$0.116$\pm$0.002 & arcsec/yr \\
$\pi$                      &    0.142$\pm$0.038 & arcsec \\
$d_{trig}$                 &      7.0$\pm$1.9   & pc \\
$d_{photom,1}$\tablefootmark{a} & 5.0$\pm$1.2   & pc \\
$d_{photom,2}$\tablefootmark{b} & 6.0/6.6/6.0/6.5/8.4 & pc \\
$d_{adopted}$              &     $\approx$5-8      & pc \\
$v_{tan}$                  &     $\approx$3-5      & km/s \\
\hline   
\end{tabular}
\tablefoot{
\tablefoottext{a}{Based on comparison with absolute $J$ and $w2$
magnitudes of LP~944-20}
\tablefoottext{b}{Applying the spectral type/absolute magnitude relations
of Dupuy \& Liu~(\cite{dupuy12}) to spectral types L0/M9.5/M9/M8.5/M8}
}
\end{table}


\section{Photometric classification and distance}
\label{Sect_phot}

For comparison of the colours of WISE~J0720$-$0846 with
those of other objects (bottom part of Table~\ref{table_colours}), 
first we 
selected the nearest known (with distances based on
trigonometric
parallaxes given in the top row of Table~\ref{table_colours}) 
representatives with spectral types between M8 and L1
not known to be part of close binaries, 
using the compilations of 
Kirkpatrick et al.~(\cite{kirkpatrick12}) and Gelino et al.~(\cite{gelino12}).
Note that the two nearby L0 and L1 dwarfs in Table~\ref{table_colours}
are fainter than the
limiting magnitude ($w2$$<$10) applied in our search (Sect.~\ref{Sect_search}).
For the M8 dwarf, VB~10, we found two counterparts observed at different epochs
in the CMC with rather different $r'$ magnitudes (15.549 and 16.410) and
used the average in computing the $r'$$-$$J$ colour.
Only the L1 comparison object has an $r'$ magnitude in the
Sloan Digital Sky Survey (SDSS), data release 7 
(Abazajian et al.~\cite{abazajian09}) that can be compared with CMC
photometry.
The $r'$$-$$J$ colours are just given for completeness, but not
discussed further as the CMC errors are unknown, and as there may
be systematic errors at the faint end of the CMC.

\begin{figure}
   \centering
   \includegraphics[width=8.5cm]{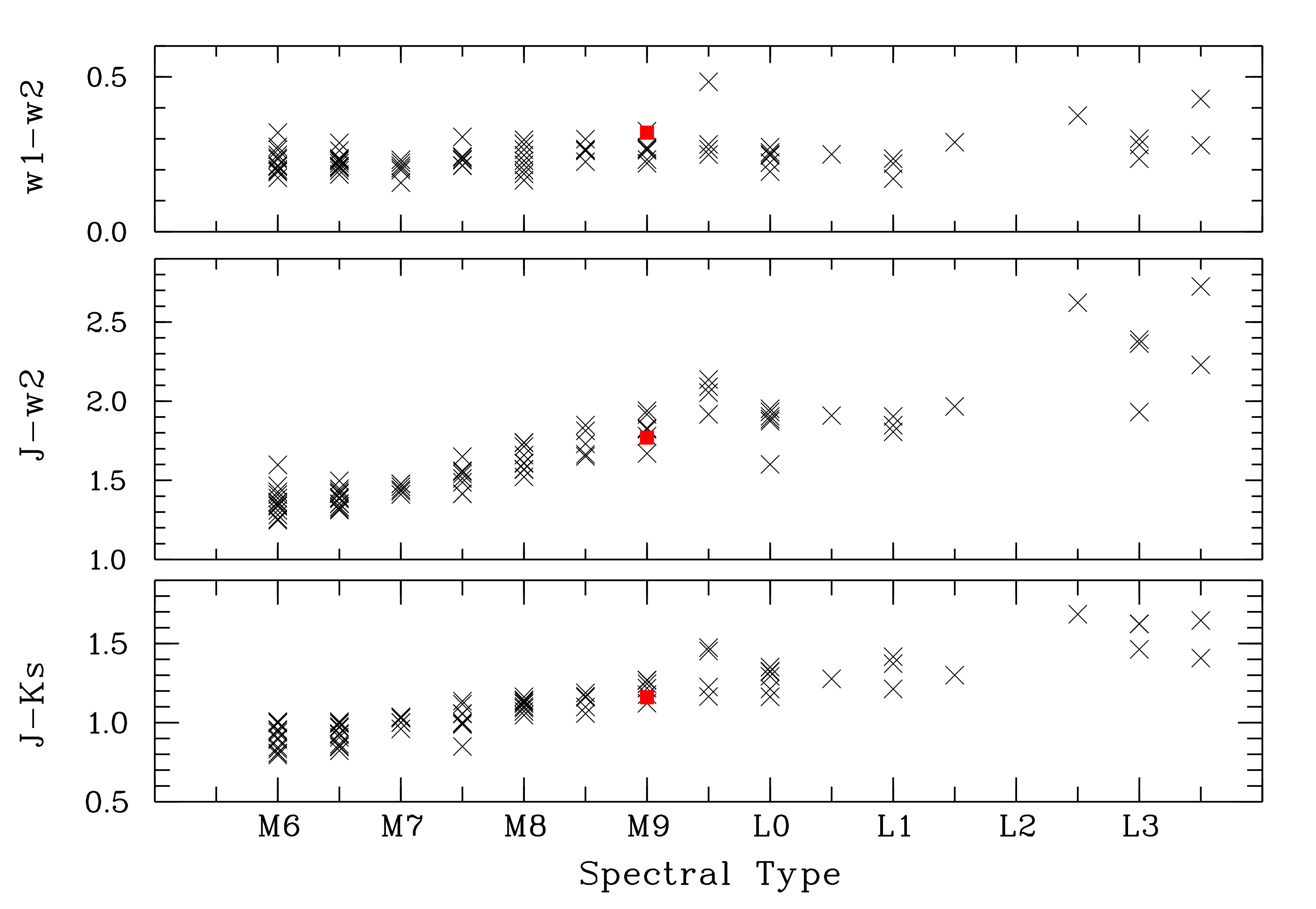}
      \caption{Colours of single M6-L3.5 dwarfs (crosses) from 
              Dupuy \& Liu~(\cite{dupuy12}) compared to those of
              WISE~J0720$-$0846 (red squares).
               }
         \label{fig3_colours}
   \end{figure}

WISE~J0720$-$0846 appears to be most similar in its colours
(and also apparent magnitudes listed in the second and third rows of
Table~\ref{table_colours}) to the M9 brown dwarf, LP~944-20.
Except for $J$$-$$w2$, all colours of these two objects
agree within 0.06~mag.
The combination of 
colours $I$$-$$J$$\approx$3.2 and $J$$-$$K_s$$\approx$1.2
is also typical of M9 dwarfs as shown in 
Fig.~2 of Phan-Bao~(\cite{phanbao08}).
However, there is an expected spread in the colour indices, and so
the new object also resembles the L1 ($J$$-$$H$ and $J$$-$$w2$), M8 ($J$$-$$H$) 
and M8.5 ($w1$$-$$w2$) dwarfs. 
The comparison of three of the colours with those of a larger
sample of objects listed in Dupuy \& Liu~(\cite{dupuy12}) is shown in
Fig.~\ref{fig3_colours}. Known close binaries, young
and overluminous objects, as well as subdwarfs, were excluded here.
Based on the comparison of colours in 
Table~\ref{table_colours} and Fig.~\ref{fig3_colours},
we tentatively classify WISE~J0720$-$0846 as an M9$\pm$1 dwarf.

The absolute magnitudes of the comparison objects (4th and 5th row of
Table~\ref{table_colours})
show a clear trend towards fainter absolute magnitudes with later spectral 
types, however this trend is interrupted by the M9 brown dwarf LP~944-20 that
is even fainter than the L1 dwarf. As LP~944-20 is apparently an outlier 
(see also, e.g., Bihain et al.~\cite{bihain10}),
but exhibits colours very similar to WISE~J0720$-$0846, we computed
two different photometric distances. One, $d_{photom,1}$ 
in Table~\ref{table_pmplx}, is the average photometric distance
obtained from the comparison of the apparent $J$ and $w2$ magnitudes
of WISE~J0720$-$0846 with the corresponding absolute magnitudes of
LP~944-20, assuming an uncertainty in the absolute magnitude of 0.5~mag.
For a second estimate of the photometric distance ($d_{photom,2}$),
we used the spectral type/absolute magnitude relations (for $M_J$ and
$M_{w2}$) provided by Dupuy \& Liu~(\cite{dupuy12}). They rely on
a larger number of typically more distant objects with measured
trigonometric parallaxes. The five values of $d_{photom,2}$ correspond to
the distances as computed for types from L0 to M8.
The mean value of $d_{photom,2}$ is 6.7$\pm$0.5~pc. The small formal
error shows that the systematic differences between the five sub-types are 
mostly smaller than the uncertainty of the absolute $J$ and $w2$ magnitudes of 
M9 dwarfs of about 0.3~mag in Dupuy \& Liu (\cite{dupuy12}). If only 
their data for M9 dwarfs are used, the result is 6.0$\pm$0.9 pc.
Although the distances obtained from the trigonometric parallax
and from photometry formally agree within the errors, the photometric 
distance tends to be smaller. 
The adopted distance of $\approx$5-8~pc is the range that is covered 
by all of our trigonometric and photometric estimates.

\begin{table}
\caption{Comparison with magnitudes/colours of nearest M/L dwarfs.} 
\label{table_colours}      
\centering                          
\begin{tabular}{lrrrrrr}        
\hline\hline                 
param. & M8\tablefootmark{a} & M8.5\tablefootmark{b} & Target & M9\tablefootmark{c} & L0\tablefootmark{d} & L1\tablefootmark{e}  \\
\hline
$d$ [pc]     &  6.1   &  4.04   &        &   5.0 &     27 &  14.5 \\
$J$          & 9.91   &  9.54   &  10.63 & 10.73 &  14.00 & 12.76 \\
$w2$         & 8.25   &  7.81   &   8.86 &  8.81 &  12.09 & 10.95 \\
$M_J$        & 11.0   &  11.5   &        &  12.2 &   11.8 &  12.0 \\
$M_{w2}$     &  9.3   &   9.8   &        &  10.3 &    9.9 &  10.1 \\
\hline
$r'$$-$$J$   & 6.07   &         &   6.22 &       &        &  6.95 \\
$I$$-$$J$    &        &  3.13   &   3.18 &  3.24 &        &       \\
$J$$-$$H$    & 0.68   &  0.63   &   0.71 &  0.71 &   0.79 &  0.72 \\
$J$$-$$K_s$  & 1.14   &  1.09   &   1.16 &  1.18 &   1.32 &  1.21 \\
$J$$-$$w2$   & 1.66   &  1.72   &   1.77 &  1.92 &   1.90 &  1.81 \\
$w1$$-$$w2$  & 0.22   &  0.29   &   0.31 &  0.32 &   0.25 &  0.24 \\
$w2$$-$$w3$  & 0.17   &  0.35   &   0.58 &  0.54 &$-$0.04 &  0.42 \\
\hline
\end{tabular}
\tablefoot{
The $r'$ magnitudes are from CMC and SDSS, $I$ from DENIS, $J$$H$$K_s$ from 
2MASS, and $w1$, $w2$, and $w3$ magnitudes are from the WISE all-sky survey. 
Object names, trigonometric parallaxes, and spectral type references:
\tablefoottext{a}{VB~10,
164.3$\pm$3.5~mas (Tinney~\cite{tinney96}),
classified as M8Ve: in Henry et al.~(\cite{henry04})}
\tablefoottext{b}{DENIS-P~J104814.7$-$395606.1,
247.71$\pm$1.55~mas (Jao et al.~\cite{jao05}),
originally classified as M9 by Delfosse et al.~(\cite{delfosse01}),
reclassified as M8.5 by Henry et al.~(\cite{henry04})}
\tablefoottext{c}{LP~944-20,
201.4$\pm$4.2~mas (Tinney~\cite{tinney96}),
originally typed as $>$M9 by Tinney \& Reid~(\cite{tinney98b}),
reclassified as M9 by Henry et al.~(\cite{henry04})}
\tablefoottext{d}{2MASP~J0345432$+$254023,
37.1$\pm$0.5~mas (Dahn et al.~\cite{dahn02}),
optical type L0 (Kirkpatrick et al.~\cite{kirkpatrick99}), 
NIR type L1$\pm$1 (Knapp et al.~\cite{knapp04})}
\tablefoottext{e}{2MASSW~J1439284$+$192915,
69.6$\pm$0.5~mas (Dahn et al.~\cite{dahn02}),
optical type L1 (Kirkpatrick et al.~\cite{kirkpatrick99})}
}
\end{table}

\section{Conclusions}
\label{Sect_concl}

The new candidate for the 8~pc sample is probably one of the 
nearest late-M or early-L dwarfs. It has a 
relatively small proper motion,
but the resulting small tangential velocity is not unusual for
a member of the Galactic thin disk population.
Given its bright magnitude and red colour in DENIS data, 
it is unclear why this M9 dwarf candidate
was not already found in the Galactic plane survey of 
Phan-Bao et al.~(\cite{phanbao08}).
Classification spectroscopy is needed to confirm the photometrically
estimated spectral type, and medium-resolution spectroscopy will
allow for the lithium test to clarify if this is a brown dwarf.
Future observations will also show whether the larger distance derived from
the preliminary trigonometric parallax compared to most of the photometric
distance estimates may indicate that WISE~J0720$-$0846 is,
in fact, a close binary. So this is
a very promising new target for high-resolution imaging, but also
for trigonometric
parallax and planet-search programs, suitable for observatories in the 
Northern and Southern hemispheres.

\smallskip

\textit{Note added in proof}\footnotesize{:
The new trigonometric parallax measured for LP~944-20 by
Dieterich et al. (2014, AJ, in press, arXiv:1312.1736)
leads to $d_{photom,1}$$=$6.3$\pm$1.5~pc for WISE~J0720$-$0846,
which is in better agreement with $d_{photom,2}$ and $d_{trig}$
and yields $d_{adopted}$$\approx$6-8~pc.}


\begin{acknowledgements}
This research has made use of the
NASA/IPAC Infrared Science Archive, which is operated by the Jet Propulsion
Laboratory, California Institute of Technology, under contract with the
National Aeronautics and Space Administration,
of data products from WISE,
which is a joint project of the University of California,
Los Angeles, and the Jet Propulsion Laboratory/California Institute of
Technology, funded by the National Aeronautics and Space Administration,
and from 2MASS.
We have also extensively used the SIMBAD database and the VizieR catalogue
access tool operated at the CDS,
including the CDS xMatch service. We thank Jesper Storm and Axel Schwope 
for helpful discussion and the anonymous referee for her/his 
comments and suggestions, which helped us improve this paper.
\end{acknowledgements}


\end{document}